\documentclass[twocolumn,prb,amsfonts,amsmath,amssymb,floatfix,superscriptaddress]{revtex4-2} 
\usepackage{color}
\usepackage{hhline}
\usepackage{mathrsfs}
\usepackage{graphicx}
\usepackage{dcolumn}
\usepackage{bm}
\usepackage{multirow}
\usepackage{booktabs}
\usepackage{afterpage}
\usepackage{amsmath}
\usepackage[T1]{fontenc}
\usepackage{ulem}
\graphicspath{{fig/}}

\begin{document}

\title{Theoretical study of the crystal structure of the bilayer nickel oxychloride Sr$_3$Ni$_2$O$_5$Cl$_2$ and analysis of possible unconventional superconductivity}

\author{Masayuki Ochi}
\thanks{These authors contributed equally to this work.}
\affiliation{Forefront Research Center, Osaka University, 1-1 Machikaneyama-cho, Toyonaka, Osaka 560-0043, Japan}
\affiliation{Department of Physics, Osaka University, 1-1 Machikaneyama-cho, Toyonaka, Osaka 560-0043, Japan}

\author{Hirofumi Sakakibara}
\thanks{These authors contributed equally to this work.}
\affiliation{Advanced Mechanical and Electronic System Research Center (AMES), Faculty of Engineering, Tottori University, 4-10 Koyama-cho, Tottori, Tottori 680-8552, Japan}

\author{Hidetomo Usui}
\affiliation{Department of Applied Physics, Shimane University, 1060 Nishikawatsu-cho, Matsue, Shimane 690-8504, Japan}

\author{Kazuhiko Kuroki}
\email{kuroki@phys.sci.osaka-u.ac.jp}
\affiliation{Department of Physics, Osaka University, 1-1 Machikaneyama-cho, Toyonaka, Osaka 560-0043, Japan}

\date{\today}
\begin{abstract}
The discovery of superconductivity under high pressure with $T_c$ exceeding 80 K in a bilayer nickelate La$_3$Ni$_2$O$_7$ has led to a strong desire to realize similar high $T_c$ phenomena at ambient pressure.
As one possible path toward realizing superconductivity at ambient pressure, we here propose to consider Sr$_3$Ni$_2$O$_5$Cl$_2$ as a possible candidate.
In this study, we theoretically investigate the electronic structure of Sr$_3$Ni$_2$O$_5$Cl$_2$ and its structural stability.
Our phonon calculation shows that this compound with the $I4/mmm$ tetragonal structure is dynamically stable even at ambient pressure.
The characteristic crystal field in this compound lowers the Ni-$d_{3z^2-r^2}$ orbital energy, by which the Ni-$d_{3z^2-r^2}$ orbital becomes rather closer to the half-filling in Sr$_3$Ni$_2$O$_5$Cl$_2$ than La$_3$Ni$_2$O$_7$. As a result, we find that superconductivity is enhanced even though a relatively strong orbital hybridization between the $t_{2g}$ and $e_g$ orbitals is somewhat detrimental for superconductivity. We also check the formation enthalpy, which shows that the high-pressure synthesis can be a good way to actually produce Sr$_3$Ni$_2$O$_5$Cl$_2$.
We find that Sr$_3$Ni$_2$O$_5$Cl$_2$ is a promising new candidate of bilayer-nickelate superconductors, which can possess even higher $T_c$ than pressurized La$_3$Ni$_2$O$_7$, at ambient pressure.
\end{abstract}

\maketitle

\section{Introduction}

\begin{figure}
\begin{center}
\includegraphics[width=8.5 cm]{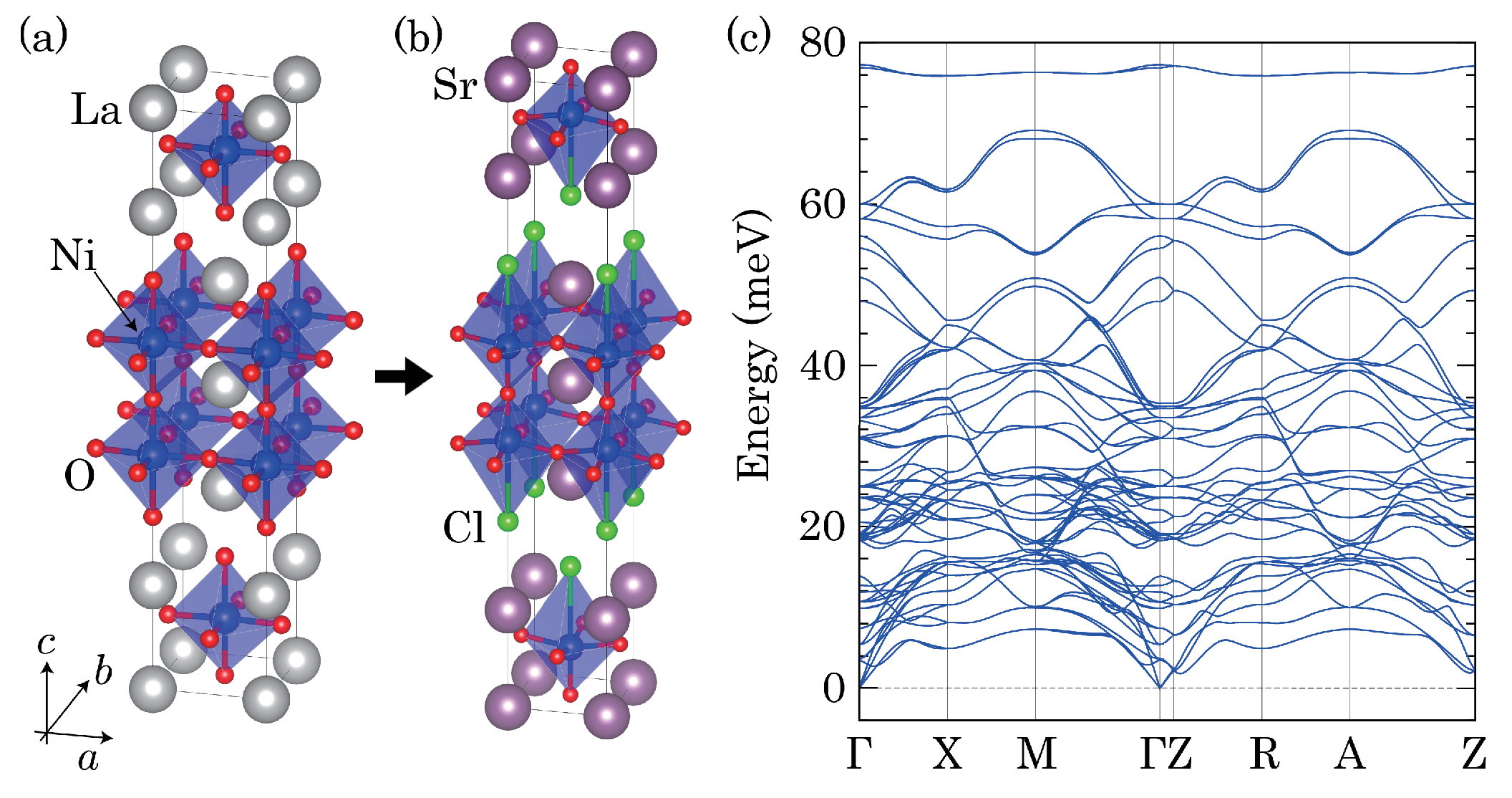}
\caption{(a) Crystal structure of La$_3$Ni$_2$O$_7$ at high pressure. (b) Crystal structure and (c) phonon band dispersion for Sr$_3$Ni$_2$O$_5$Cl$_2$ without applying pressure. Panels (b) and (c) are shown for the space group of $I4/mmm$.
Crystal structures in this paper were depicted using the VESTA software~\cite{VESTA}. Phonon dispersions in this paper were depicted in the Brillouin zone of the conventional tetragonal lattice.}
\label{fig:3252_phonon}
\end{center}
\end{figure}

Discovery of superconductivity with a maximum $T_c$ of above 80 K in a bilayer Ruddlesden-Popper nickelate La$_3$Ni$_2$O$_7$ [Fig.~\ref{fig:3252_phonon}(a)] under pressure has sparked enormous interest in the field of condensed matter physics~\cite{Sun_Huo}. Although under pressure, $T_c$ exceeding the liquid nitrogen boiling temperature, and also the origin of superconductivity likely being unconventional~\cite{Ouyang_Gao,Li_Cao,Yi_Meng}, have led to a huge amount of experimental and theoretical studies after the discovery~\cite{Sun_Huo,Zhang_Lin_2,Yang_Wang,Lechermann_Gondolf,Sakakibara_Kitamine,Gu_Le,Shen_Qin,Christiansson_Petocchi,Liu_Huo,Wu_Luo,Cao_Yang,Hou_Yang,Liu_Mei,Zhang_Su,Lu_Pan_15,Zhang_Lin_16,Oh_Zhang,Liao_Chen,Qu_Qu_19,Yang_Zhang,Jiang_Wang,Zhang_Lin_22,Tian_Chen,Jiang_Hou,Luo_Lv,Yang_Sun,Zhang_Pei_29,Sakakibara_Ochi,Geisler_Hamlin,Rhodes_Wahl,Wang_Wang_39,Kaneko_Sakakibara,Lu_Pan_41,Ryee_Witt,Li_Zhang,Zhu_Peng,Zhang_Pei_49,Wang_Wang_50,Wang_Li,Ouyang_Wang,Yuan_Elghandour,Zhou_Guo,Qu_Qu_56,Chen_Liu,Li_Chen,Sui_Han,Kakoi_Kaneko,Chen_Zhang,Puphal_Reiss,Li_Guo,Wang_Chen,Kakoi_Oi,Dong_Huo,Xie_Huo,Chen_Choi,Wang_Jiang,Tian_Ma,Wang_Ouyang,Dan_Zhou,LaBollita_Kapeghian,Zhang_Lin_83,Yang_Jiang,Lu_Pan_85,Abadi_Xu,Khasanov_Hicken,Li_Cao,Yi_Meng,Wu_Yang,Ouyang_Gao,Li_Ma,Meng_Yang,Wang_Zhou,Xu_Chen,Du_Li,Nagata_Sakurai,Wang_Wang_111,Ueki_Sakurai}. In fact, the possibility of superconductivity in this material was discussed theoretically even before its experimental discovery by some of the previous authors~\cite{Nakata}, but one viewpoint that was missing at that time was the precise crystal structure, namely, although a tetragonal symmetry was simply assumed in Ref.~\onlinecite{Nakata}, the actual structural symmetry is orthorhombic at ambient pressure.  There are now accumulating studies suggesting that tetragonal symmetry of the crystal structure may be important for the occurrence of superconductivity in La$_3$Ni$_2$O$_7$~\cite{Geisler_Hamlin,Wang_Li,Wang_Wang_50}, which is accomplished by applying pressure.
One speculation regarding the reason for this can be that orthorhombicity favors density waves, which suppresses superconductivity.
Interestingly, also for the trilayer compound La$_4$Ni$_3$O$_{10}$, the superconducting phase appears by applying a sufficient pressure and the tetragonal structure is realized~\cite{Sakakibara_Ochi,Li_Zhang,Zhu_Peng,Zhang_Pei_49,Li_Chen}.
Then, a natural motivation arises for searching materials that have tetragonal symmetry at ambient pressure, and also have electronic properties similar to that of La$_3$Ni$_2$O$_7$ at high pressures.

One path toward realizing tetragonal symmetry at ambient pressure is to replace the rare earth (La) atom with an element with larger ion radius, which however is not possible within lanthanoids since La has the largest ion radius~\cite{Zhang_Lin_22,Geisler_Hamlin}. Hence, there is a proposal for using actinium instead~\cite{Rhodes_Wahl,Wu_Yang}, which however is strongly radioactive. Alkaline-earth elements such as Sr and Ba~\cite{Rhodes_Wahl} have larger ion radius than La, but the valence is $2+$ rather than $3+$, which results in a nickel valence of $4+$, which is not typical. These considerations have led us to consider the possibility of Sr$_3$Ni$_2$O$_5$Cl$_2$ (Fig.~\ref{fig:3252_phonon}(b)) as a possible candidate for our purpose, where the two chlorines replace the outer apical oxygens to reduce the nickel valence. The nickel valence is $3+$ as compared to $2.5+$ in La$_3$Ni$_2$O$_7$, but an expectation is that the $d_{3z^2-r^2}$ band filling, which has been shown to be crucial for superconductivity in La$_3$Ni$_2$O$_7$, could be similar because the energy level offset $\Delta E$ between $d_{3z^2-r^2}$ and $d_{x^2-y^2}$ is expected to be larger in Sr$_3$Ni$_2$O$_5$Cl$_2$ due to the outer apical oxygens being replaced by chlorine.   Although this material has not been synthesized to our knowledge, its iron and cobalt analogs Sr$_3$Fe$_2$O$_5$Cl$_2$~\cite{Leib1984} and Sr$_3$Co$_2$O$_5$Cl$_2$~\cite{Cava} have been synthesized in the past, and the crystal structure is known to have tetragonal symmetry. 
It is noteworthy that oxychlorides have also been investigated in the study of cuprate superconductors such as (Ca,Na)$_2$CuO$_2$Cl$_2$~\cite{Hiroi_cuprate_ocl}, where the crystal field of the transition metal atom and the carrier concentration are changed via anion substitution.
Recently, another mixed-anion nickelate La$_2$$Ae$Ni$_2$O$_6$F has been proposed as a material having tetragonal symmetry at ambient ($Ae=$ Ba) or low ($Ae=$ Sr) pressure~\cite{Wu_Yang}, but our proposal is different in that the inner apical oxygen, which we believe to be playing a crucial role in the occurrence of superconductivity in La$_3$Ni$_2$O$_7$ is maintained, while outer apical oxygens are replaced by halogens.

In this study, we theoretically investigate the electronic structure of Sr$_3$Ni$_2$O$_5$Cl$_2$ and its structural stability.
Our phonon calculation shows that this compound with the $I4/mmm$ tetragonal structure is dynamically stable, i.e., does not have imaginary-frequency phonon modes, even at ambient pressure.
The Ni-$d_{3z^2-r^2}$ orbital energy is pushed down by the crystal field where half of the apical O$^{2-}$ ions are replaced with a Cl$^{-}$ ion. By this effect, the Ni-$d_{3z^2-r^2}$ orbital becomes rather closer to the half-filling in Sr$_3$Ni$_2$O$_5$Cl$_2$ than La$_3$Ni$_2$O$_7$ whereas the stoichiometric Ni-$3d$ band filling are decreased in
Sr$_3$Ni$_2$O$_5$Cl$_2$ ($d^7$) compared with La$_3$Ni$_2$O$_7$ ($d^{7.5}$).
We also investigate superconductivity in this compound with the multi-orbital Hubbard model constructed from our first-principles calculation, using the fluctuation exchange (FLEX) approximation combined with the linearized Eliashberg equation.
While orbital hybridization between the $t_{2g}$ and $e_g$ orbitals, which is enhanced in this characteristic crystal field, is detrimental for superconductivity,
a nearly half-filled Ni-$d_{3z^2-r^2}$ orbital enhances superconductivity.
In total, we find that Sr$_3$Ni$_2$O$_5$Cl$_2$ is a promising new candidate of bilayer-nickelate superconductors, which can possess even higher $T_c$ than pressurized La$_3$Ni$_2$O$_7$, at ambient pressure.
We also check the formation enthalpy for synthesis of this compound. Our calculation shows that the high-pressure synthesis can be a good way to experimentally produce Sr$_3$Ni$_2$O$_5$Cl$_2$.

\section{Methods}

For first-principles calculation, we used the PBEsol exchange-correlation functional~\cite{PBEsol} and the projector augmented wave (PAW) method as implemented in Vienna {\it ab initio} Simulation Package (VASP)~\cite{paw,vasp1,vasp2,vasp3,vasp4}. 
Core-electron states in PAW potentials were [Kr]$4d^{10}$, [Ar]$3d^{10}$, [Ar], [Ne], and [He] for La, Sr, Ni, Cl, and O, respectively.
We used a plane-wave cutoff energy of 600 eV for Kohn-Sham orbitals without including the spin-orbit coupling for simplicity.
A $12\times 12\times 12$ ${\bm k}$-mesh was used for self-consistent-field calculations.

We performed structural optimization until the Hellmann-Feynman force becomes less than 0.01 eV \AA$^{-1}$ for each atom.
To check the stability of the tetragonal ($I4/mmm$) phase, we calculated the phonon dispersion for several materials.
For this purpose, we used the finite displacement method as implemented in the \textsc{Phonopy}~\cite{phonopy1,phonopy2} software in combination with VASP.
We used a $4\times 4\times 1$ ${\bm q}$-mesh for a conventional tetragonal unit cell.
For a $4\times 4\times 1$ supercell used for finite-displacement calculations, we used a $3\times 3\times 2$ ${\bm k}$-mesh.

To discuss superconductivity, we constructed a model Hamiltonian of Ni-$3d$ orbitals in the following way.
We extracted five $3d$-like Wannier orbitals per Ni site using \textsc{Wannier90} software~\cite{Wannier1,Wannier2,Wannier90}.
For this purpose, we used a $12\times 12\times 12$ ${\bm k}$-mesh for a primitive unit cell.
We constructed a model Hamiltonian by adding the on-site interaction terms to the tight-binding Hamiltonian obtained through Wannierization.
We took the on-site interactions, namely, intraorbital (interorbital) Coulomb interactions $U$ ($U'$), 
Hund's coupling $J$, and pair hopping $J'$.
We assumed the orbital-rotational symmetry, namely, we took the same value of $U$ for all 3$d$ orbitals preserving the equations $U'=U-2J$ and $J=J'$.  
We took $U=3$ eV, $J=J'=U/10=0.3$ eV and $U'=U-2J=2.4$ eV, which can be considered as typical values for 3$d$-transition-metal oxides.

We adopted FLEX~\cite{Bickers,Bickers1991} to analyze electron correlation effects for the Hubbard model as was done in Ref.~\cite{Sakakibara_Kitamine}.
We calculated the self-energy induced by the spin-fluctuation formulated as shown in the literatures~\cite{Lichtenstein,mFLEX1,mFLEX2} in a self-consistent calculation.
The real part of the self-energy at the lowest Matsubara frequency was subtracted in the same manner as Ref.~\cite{Ikeda_omega0} to maintain the band structure around the Fermi level obtained by first-principles calculation.
We used the linearized Eliashberg equation to study the possibility of superconductivity in Sr$_3$Ni$_2$O$_5$Cl$_2$ comparing with La$_3$Ni$_2$O$_7$.
The renormalized Green's functions obtained by FLEX were plugged into this equation. Note that the pairing interaction kernel in this equation was obtained from the FLEX Green's function as a purely electronic one (i.e. phonon-mediated pairing interaction was not considered), which is mainly dominated by spin fluctuations in the present case.
Since the eigenvalue $\lambda$ of the Eliashberg equation monotonically increases upon lowering the temperature $T$, and reaches unity at $T=T_c$, 
we adopted $\lambda$ calculated at a fixed temperature, $T=0.01$ eV, as a measure of superconductivity. For convenience, we will call the eigenfunction (with the largest eigenvalue) of the linearized Eliashberg equation at the lowest Matsubara frequency $i\omega$ $=$ $i\pi k_{\rm B}T$ the ``superconducting gap function''. We took a 16$\times$16$\times$4 ${\bm k}$-mesh and 2048 Matsubara frequencies for the FLEX calculation.

For calculating the formation energy, we adopted space groups taken from experimental studies: 
$Fm{\bar 3}m$ for SrO, SrCl$_2$, Ni, NiO,
$I4/mmm$ for SrCl$_2$, Sr$_2$NiO$_2$Cl$_2$~\cite{Tsujimoto},
$Cmcm$ for SrNiO$_2$~\cite{Pausch1976}, and
$P6_3/mmc$ for SrNiO$_3$~\cite{Takeda1972}.
An isolated O$_2$ molecule was calculated in a 15 {\AA} $\times$ 15 {\AA} $\times$ 15 {\AA} cell considering the spin-triplet ground state with the spin-polarized calculation.

\section{Results and Discussions}

\subsection{Electronic and phonon band structures for Sr$_3$Ni$_2$O$_5$Cl$_2$\label{sec:FP}}

As described in the Introduction, a key idea in this study is to employ a mixed-anion strategy~\cite{mixed_anion_review} for stabilizing the tetragonal $I4/mmm$ phase without applying the pressure.
In fact, for Sr$_3$Ni$_2$O$_5$Cl$_2$, while the NiO$_2$ plane is somewhat buckled; the equatorial Ni-O-Ni angle is 162$^{\circ}$ as shown in Fig.~\ref{fig:3252_phonon}(b),
we found that the tetragonal structure ($a= 3.779$ \AA, $c=24.532$ \AA) is dynamically stable as shown by our phonon calculation (Fig.~\ref{fig:3252_phonon}(c)) where no imaginary-frequency modes appear without applying the pressure.
This is in sharp contrast to the case of La$_3$Ni$_2$O$_7$, where the tetragonal structure is unstable at ambient pressure and high pressure is required to stabilize it~\cite{Geisler_Hamlin,Wang_Li,Wang_Wang_50}.
Our phonon calculations for La$_3$Ni$_2$O$_7$ also support this feature as shown in Fig.~\ref{fig:327_phonon}, where the imaginary-frequency modes appear for 0 GPa while they are absent for 20 GPa.
Although we also checked the stability of the tetragonal structure for Sr$_3$Ni$_2$O$_4$Cl$_3$, where all the apical oxygens were replaced with chlorines, we found that the tetragonal structure was unstable for this case as shown in Fig.~\ref{fig:3243}.
We consider that one important difference among these compounds is the average valence number of the nickel atom:  Ni$^{2.5+}$ ($d^{7.5}$) for La$_3$Ni$_2$O$_7$ and Sr$_3$Ni$_2$O$_4$Cl$_3$ while
 Ni$^{3+}$ ($d^7$) for Sr$_3$Ni$_2$O$_5$Cl$_2$.
Here, the size of the nickel ion becomes smaller in Sr$_3$Ni$_2$O$_5$Cl$_2$, which in general suppresses the rotational and tilting instability of octahedra in Ruddlesden-Popper phase materials.

\begin{figure*}
\begin{center}
\includegraphics[width=15 cm]{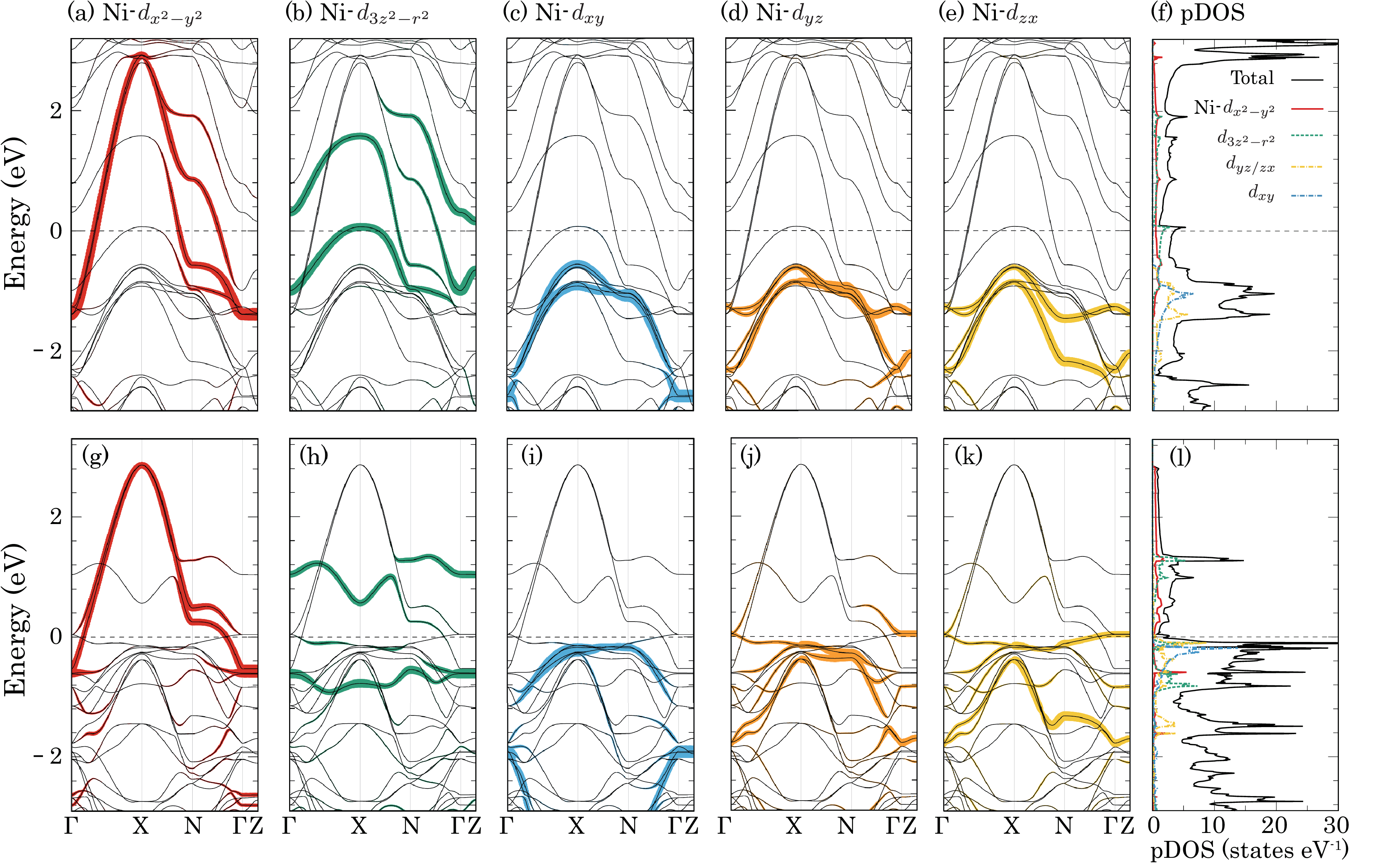}
\caption{Electronic band dispersion with a colored weight of Ni- (a) $d_{x^2-y^2}$, (b) $d_{3z^2-r^2}$, (c) $d_{xy}$, (d) $d_{yz}$, and (e) $d_{zx}$ orbitals, respectively, and (f) pDOS for La$_3$Ni$_2$O$_7$ under pressure of 20 GPa. (g)--(l) Same plots for Sr$_3$Ni$_2$O$_5$Cl$_2$ without applying the pressure.}
\label{fig:bandDOS}
\end{center}
\end{figure*}

It is noteworthy that the mixed-anion strategy drastically changes the crystal field of a nickel ion.
Figure~\ref{fig:bandDOS} presents the electronic band structures and partial density of states (pDOS) for La$_3$Ni$_2$O$_7$ under pressure of 20 GPa and Sr$_3$Ni$_2$O$_5$Cl$_2$ without applying the pressure shown with an orbital weight of each Ni-$d$ orbital. We summarize important differences between two materials from the viewpoint of the crystal field as follows.

As mentioned above, the average valence number of the nickel atom is different: Ni$^{2.5+}$ ($d^{7.5}$) for La$_3$Ni$_2$O$_7$ and Ni$^{3+}$ ($d^7$) for Sr$_3$Ni$_2$O$_5$Cl$_2$.
Thus, the number of $d$ electrons for Sr$_3$Ni$_2$O$_5$Cl$_2$ is less than that for La$_3$Ni$_2$O$_7$.
Nevertheless, the Ni-$d_{3z^2-r^2}$ orbital is rather closer to the half-filling in Sr$_3$Ni$_2$O$_5$Cl$_2$ as shown in Figs.~\ref{fig:bandDOS}(b)(h).
This is because of the characteristic crystal field for the nickel atom in Sr$_3$Ni$_2$O$_5$Cl$_2$, where a half of apical O$^{2-}$ ions is replaced with a Cl$^{-}$ ion, which pushes down the orbital energy of the Ni-$d_{3z^2-r^2}$ orbital. As a result, the number of $d_{x^2-y^2}$ electrons is decreased in Sr$_3$Ni$_2$O$_5$Cl$_2$.
On the other hand, Ni-$d_{xy/yz/zx}$ orbitals, which we hereafter call the $t_{2g}$ orbitals while it is not a rigorous term for this symmetry, raise their energy levels relative to the Ni-$d_{3z^2-r^2}$ orbital. This is because the present crystal field strongly pushes down the Ni-$d_{3z^2-r^2}$ orbital level as we have stated above.
Consequently, the strong DOS peak of the $t_{2g}$ orbitals lies just below the Fermi energy.
In addition, the crystal field is strongly asymmetric with respect to the $xy$ plane (i.e., $+z$ and $-z$ are strongly inequivalent), which allows a sizable hybridization among $e_g$ ($d_{x^2-y^2}$ and $d_{3z^2-r^2}$) and $t_{2g}$ orbitals. 
In the next section, we will discuss how these characteristics can affect superconductivity in this system.

\subsection{Model construction}
\begin{figure}
\begin{center}
\includegraphics[width=8cm]{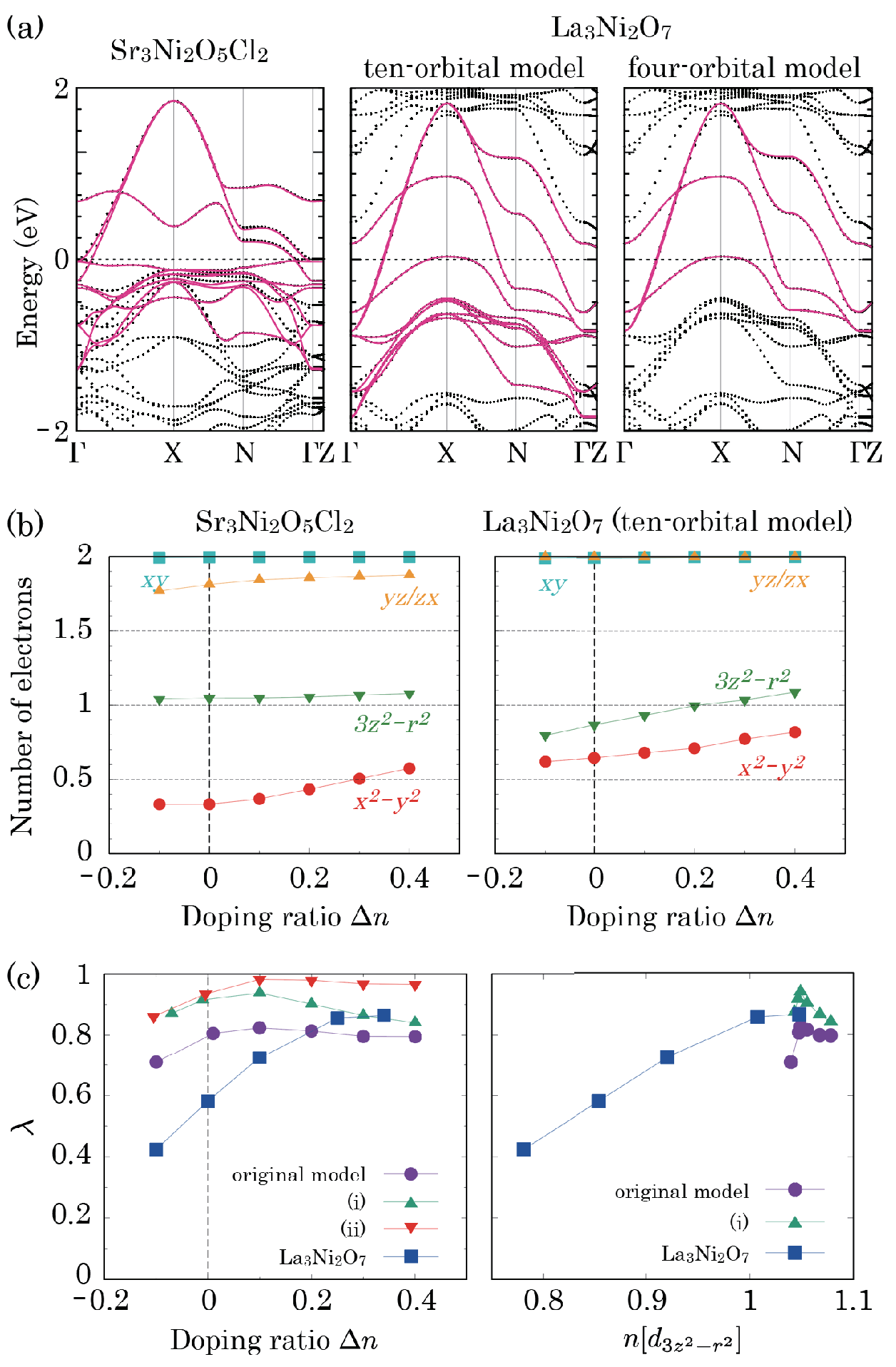}
\caption{(a) Band structure of the ten-orbital model (pink solid lines) superposed on that obtained from first-principles calculation (black dotted lines) for Sr$_3$Ni$_2$O$_5$Cl$_2$ (left) and those for the ten-orbital model (middle) and the four-orbital model (right) of La$_3$Ni$_2$O$_7$,
(b) the number of electrons per Ni for each orbital as a function of doping ratio $\Delta n$ for Sr$_3$Ni$_2$O$_5$Cl$_2$ (left) and La$_3$Ni$_2$O$_7$ (right),
(c) the eigenvalue $\lambda$ of the linearized Eliashberg equation at $T=0.01$ eV as a function of
$\Delta n$ (left) and $n[d_{3z^2-r^2}]$ (right),
where the case of stoichiometry is indicated by arrows.
In panel (c), the legend (i) indicates the case that the $t_{2g}-e_g$ interorbital interactions are turned off, and (ii) all the interorbital interactions are turned off. The legend ``La$_3$Ni$_2$O$_7$'' in panel (c) indicates the result of four-orbital-model calculation.
}
\label{fig:flex_lambda}
\end{center}
\end{figure}

In this and the next sections, we compared Sr$_3$Ni$_2$O$_5$Cl$_2$ without applying the pressure and La$_3$Ni$_2$O$_7$ under pressure of 20 GPa, both of which have the $I4/mmm$ tetragonal structure in our calculation.
We constructed a ten-orbital model consisting of five $3d$ orbitals centered at two Ni sites per unit cell.
Figure~\ref{fig:flex_lambda}(a) shows superposed band structures given by first-principles calculation and Wannier interpolation, where precise fitting around the Fermi level was achieved.
Important parameter values extracted by Wannierization are given in Table~\ref{tab:hopping}, compared with those of four-orbital models consisting only of $e_g$ orbitals. 
The smaller differences in extracted parameters between the ten- and four-orbital models for La$_3$Ni$_2$O$_7$ originate from 
the fact that the $e_g$ and $t_{2g}$ bands are well separated compared with Sr$_3$Ni$_2$O$_5$Cl$_2$, where the $t_{2g}$ and $e_g$ bands are rather entangled, as shown in Fig.~\ref{fig:bandDOS}.
We found a relatively large hopping amplitude between $t_{2g}$ and $e_g$ orbitals: e.g., the nearest-neighbor intralayer hopping along the $x$ direction between the $d_{x^2-y^2}$ and $d_{xz}$ orbitals
and that between the $d_{3z^2-r^2}$ and $d_{xz}$ orbitals are $-0.296$ and $0.199$ eV, respectively.
This is due to the strongly asymmetric crystal field for a NiO$_5$Cl octahedron as mentioned in the previous section.
It is also noteworthy that the relatively strong buckling of NiO$_2$ plane in Sr$_3$Ni$_2$O$_5$Cl$_2$ seems to enhance he vertical interlayer hopping $t_{\perp}$ between the $d_{3z^2-r^2}$ orbitals while the nearest-neighbor intralayer hoppings $t_{3z^2-r^2}$, $t_{x^2-y^2}$, and $t_{x^2-y^2{\text -}  3z^2-r^2}$ are somewhat suppressed as shown in Table~\ref{tab:hopping}.
For Sr$_3$Ni$_2$O$_5$Cl$_2$, we also found that the onsite energies of the $d_{xy}$ and $d_{yz/zx}$ orbitals relative to that for the $d_{3z^2-r^2}$ orbital are $-0.811$ eV and $-0.840$ eV, respectively, the size of which are comparable to $\Delta E=E_{x^2-y^2}-E_{3z^2-r^2} = 0.776$ eV listed in Table~\ref{tab:hopping}.

\begin{table}[!h]
\caption{The orbital level offset $\Delta E=E_{x^2-y^2}-E_{3z^2-r^2}$ between the $d_{x^2-y^2}$ and the $d_{3z^2-r^2}$ orbitals, the vertical interlayer hopping $t_{\perp}$ between the $d_{3z^2-r^2}$ orbitals, and the nearest-neighbor intralayer hoppings $t_{3z^2-r^2}$, $t_{x^2-y^2}$, and $t_{x^2-y^2{\text -}  3z^2-r^2}$ are displayed.
The first column indicates the chemical formulas of the calculated materials, where the number in the round brackets indicate the number of orbitals.}
\label{tab:hopping}
\scalebox{0.9}{
\begin{tabular}{c c c c c c l} \hline\hline
  (eV) &  \hspace{2pt}$\Delta E$ & \hspace{1pt}$t_{\perp}$ & \hspace{1pt}$t_{3z^2-r^2}$ & \hspace{1pt} $t_{x^2-y^2}$ & \hspace{1pt}$t_{x^2-y^2{\text -} 3z^2-r^2}$ \\\hline
 Sr$_3$Ni$_2$O$_5$Cl$_2$ (10) & \hspace{1pt}$0.776$  \hspace{2pt}& \hspace{1pt}$-0.744$ & \hspace{1pt}$0.041$ & \hspace{1pt}$-0.425$ &\hspace{1pt} $-0.086$ \\
  Sr$_3$Ni$_2$O$_5$Cl$_2$ (4) & \hspace{1pt}$0.920$  \hspace{2pt}& \hspace{1pt}$-0.786$ & \hspace{1pt}$0.037$ & \hspace{1pt}$-0.383$ &\hspace{1pt} $-0.074$ \\
La$_3$Ni$_2$O$_7$ (10) & \hspace{1pt}$0.343$  \hspace{2pt}& \hspace{1pt}$-0.660$ & \hspace{1pt}$-0.126$ & \hspace{1pt}$-0.499$ &\hspace{1pt} $-0.253$ \\
La$_3$Ni$_2$O$_7$ (4) & \hspace{1pt}$0.328$  \hspace{2pt}& \hspace{1pt}$-0.659$ & \hspace{1pt}$-0.128$ & \hspace{1pt}$-0.501$ &\hspace{1pt} $-0.253$ \\
\hline\hline
\end{tabular}
}
\end{table} 

In Fig.~\ref{fig:flex_lambda}(b), we plot the number of electrons for each orbital per Ni against the hypothetically varied total band filling, assuming a rigid band.
In this paper, the total band filling $n$ is defined as the number of electrons per spin in the primitive unit cell, e.g., $0\leq n \leq 10$ for ten-orbital models.
The doping ratio $\Delta n$ is defined as the deviation of the total band filling from its value at stoichiometry.
In Sr$_3$Ni$_2$O$_5$Cl$_2$, the number of electrons occupying the $d_{3z^2-r^2}$ band per Ni, called $n[d_{3z^2-r^2}]$ hereafter, exceeds unity (half-filled) already at stoichiometry, consistent with the observation given in Sec.~\ref{sec:FP}. Also, the $d_{yz/zx}$ orbitals are not fully filled due to the hybridization with the $e_g$ orbitals. Contrastingly, in La$_3$Ni$_2$O$_7$, the $d_{yz/zx}$ orbitals are fully filled in the whole doping range suggesting their irrelevance for superconductivity and justifying the previously proposed four-orbital model~\cite{Sakakibara_Kitamine}.

Based on the observation described in this section, we used the ten-orbital model for Sr$_3$Ni$_2$O$_5$Cl$_2$ and the four-orbital model (i.e., $e_g$ model) for La$_3$Ni$_2$O$_7$ at 20 GPa in the following analysis of superconductivity.

\subsection{FLEX analysis on superconductivity in Sr$_3$Ni$_2$O$_5$Cl$_2$}
In the left panel of Fig.~\ref{fig:flex_lambda}(c), we plot the eigenvalue $\lambda$ of the linearized Eliashberg equation against $\Delta n$.
For comparison, we also plot $\lambda$ for models where (i) the $t_{2g}-e_g$ interorbital interactions are turned off, and (ii) all the interorbital interactions are turned off. $\lambda$ is also plotted for the four-orbital model of La$_3$Ni$_2$O$_7$.
By comparing $\lambda$ calculated with the original (full) model and that calculated by the condition (i),
it can be seen that the $t_{2g}-e_g$ interorbital interactions somewhat suppress superconductivity in Sr$_3$Ni$_2$O$_5$Cl$_2$. Still, quite interestingly, the value of $\lambda$ of the full model of Sr$_3$Ni$_2$O$_5$Cl$_2$ is significantly larger than that of La$_3$Ni$_2$O$_7$ at stoichiometry, suggesting the possibility that the material possesses a potential of even higher $T_c$ than 80 K {\it at ambient pressure}. Also, a slight electron doping may result in an even higher $T_c$. 
Electron doping is expected to be possible by partially substituting Sr by La, for which we have also verified the stability of the tetragonal symmetry as shown in Fig.~\ref{fig:La10per}(b), where no imaginary-frequency phonon modes appear for La$_{0.3}$Sr$_{2.7}$Ni$_2$O$_5$Cl$_2$ in our first-principles calculation.
On the other hand, hole doping is not good for superconductivity.

\begin{figure}
\begin{center}
\includegraphics[width=9cm]{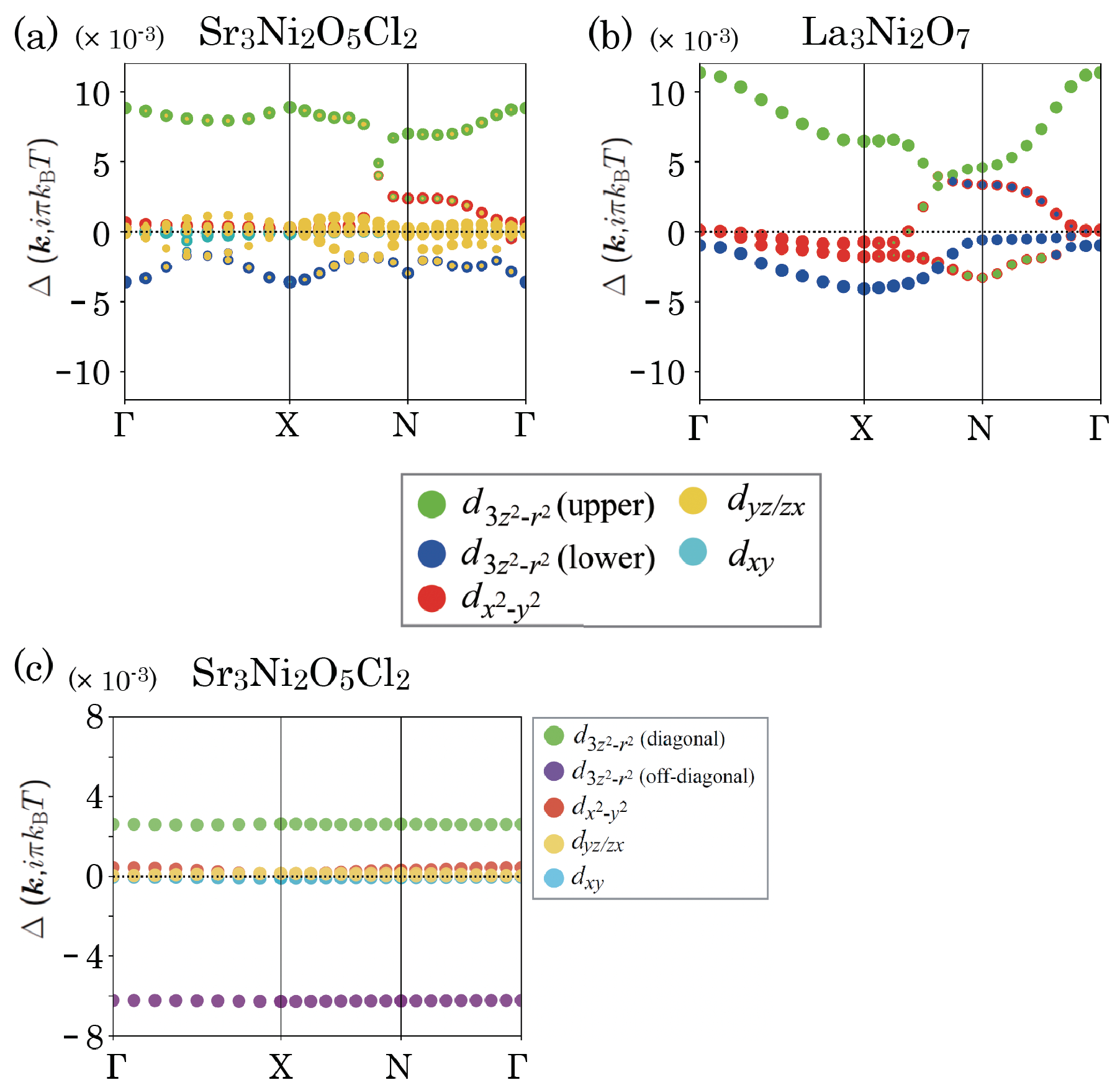}
\caption{(a) The superconducting gap function for Sr$_3$Ni$_2$O$_5$Cl$_2$ ($\Delta n = 0.01$) at $T=0.01$ eV in the band representation, compared with (b) that for the four-orbital model of La$_3$Ni$_2$O$_7$ ($\Delta n =0$).
The strength of the Wannier orbital characters is shown with the radius of the color coded circles,
where the weight of the $d_{3z^2-r^2}$ orbital is indicated by either green or blue, 
depending on whether the band energy is above or below $E=0.1$ eV.
(c) Same as (a) but in the orbital representation, where the orbital-diagonal components for all $d$ orbitals and the orbital-off-diagonal component for the $d_{3z^2-r^2}$ orbital are shown.}
\label{fig:flex_gap}
\end{center}
\end{figure}

In Fig.~\ref{fig:flex_gap}, we plot the gap function in the band representation for Sr$_3$Ni$_2$O$_5$Cl$_2$ ($\Delta n = 0.01$) and La$_3$Ni$_2$O$_7$ ($\Delta n =0$). For Sr$_3$Ni$_2$O$_5$Cl$_2$, the gap function of the $d_{3z^2-r^2}$ portion of the band is large with a basically $s\pm$-wave structure, i.e., $-$ sign in the bonding and $+$ sign in the antibonding band, as in La$_3$Ni$_2$O$_7$. The gap function of the $d_{x^2-y^2}$ portion is smaller compared to La$_3$Ni$_2$O$_7$ due to the smaller hybridization with $d_{3z^2-r^2}$.
The sign change between the bonding and antibonding $d_{3z^2-r^2}$ bands corresponds to the dominant contribution of the off-diagonal $d_{3z^2-r^2}$ component in the orbital representation, as shown in Fig.~\ref{fig:flex_gap}(c), which means that the pairing takes place mainly in the interlayer channel.

To understand the origin of the large eigenvalue around stoichiometry for Sr$_3$Ni$_2$O$_5$Cl$_2$, we plot $\lambda$ against $n[d_{3z^2-r^2}]$ in the right panel of Fig.~\ref{fig:flex_lambda}(c).
It can be seen from this figure that the reason for $\lambda$ being larger than La$_3$Ni$_2$O$_7$ is that $n[d_{3z^2-r^2}]$ for Sr$_3$Ni$_2$O$_5$Cl$_2$ is closer to unity at stoichiometry.
If we compare the two materials at the same $n[d_{3z^2-r^2}]$, $\lambda$ for Sr$_3$Ni$_2$O$_5$Cl$_2$ is suppressed compared to that of La$_3$Ni$_2$O$_7$ when $\Delta n<0$, where the Fermi level intersects the nearly flat band consisting of a mixture of $d_{yz/zx}$ and $d_{3z^2-r^2}$ orbitals. This suppression becomes small when the $t_{2g}-e_g$ interorbital interactions are turned off, suggesting that this interorbital interaction is harmful for superconductivity especially when the $t_{2g}$ orbitals strongly hybridize with the $d_{3z^2-r^2}$ orbital near the Fermi level. This is the origin of the reduction of $\lambda$ upon hole doping.

Here, we make a brief comment on the robustness of our results against the interaction parameters. We also calculated $\lambda$ for Sr$_3$Ni$_2$O$_5$Cl$_2$ at stoichiometry using the interaction parameters evaluated with the constrained random-phase approximation (cRPA) and obtained $\lambda = 0.767$. Although this value is slightly smaller than $\lambda=0.806$ shown in Fig.~\ref{fig:flex_gap}(c), it is still sufficiently higher than $\lambda$ for pressurized La$_3$Ni$_2$O$_7$ at stoichiometry. The obtained interaction parameters are shown in Appendix D. We also note that, in our previous study~\cite{Sakakibara_Kitamine}, we verified that $\lambda$ for La$_3$Ni$_2$O$_7$ is similar between two sets of interaction parameters: ($U, U', J=J'$) = (3, 2.4, 0.3) eV and those evaluated with cRPA~\cite{Christiansson_Petocchi}.

Our theoretical analysis illustrates the effect of the mixed-anion strategy of materials design on bilayer nickelate.
First of all, the tetragonal structure is stabilized in Sr$_3$Ni$_2$O$_5$Cl$_2$ as confirmed by our phonon calculation.
Second, the Ni-$d_{3z^2-r^2}$ orbital energy is pushed down by the crystal field where a half of apical O$^{2-}$ ions is replaced with a Cl$^{-}$ ion.
This results in two opposite effects on superconductivity as follows.
The Ni-$d_{3z^2-r^2}$ orbital becomes rather closer to the half-filling in Sr$_3$Ni$_2$O$_5$Cl$_2$ than La$_3$Ni$_2$O$_7$ whereas the stoichiometric Ni-$3d$ band filling are decreased in
Sr$_3$Ni$_2$O$_5$Cl$_2$ ($d^7$) compared with La$_3$Ni$_2$O$_7$ ($d^{7.5}$). The increase of the Ni-$d_{3z^2-r^2}$ band filling is beneficial for superconductivity, which was also discussed for La$_3$Ni$_2$O$_7$ in our previous study~\cite{Sakakibara_Kitamine}.
On the other hand, the lowered Ni-$d_{3z^2-r^2}$ orbital energy together with the low-symmetry crystal field enhances the hybridization between the $t_{2g}$ and $e_g$ orbitals, which is detrimental for superconductivity.
In total, we have seen that Sr$_3$Ni$_2$O$_5$Cl$_2$ is a promising new candidate of bilayer-nickelate superconductors, which can possess even higher $T_c$ than pressurized La$_3$Ni$_2$O$_7$, at ambient pressure.

\subsection{Formation energy (enthalpy)}

To examine the possibility for the synthesis of Sr$_3$Ni$_2$O$_5$Cl$_2$, we calculated formation energies for some possible chemical reactions as presented in Table~\ref{tab:form}.
Here, the formation energy is defined as the total energy of the right-hand side of the chemical equation relative to that for the left-hand side.
Huge (negative) formation energies for reactions (a)(b) mean that the left-hand sides of (a)(b) are very unstable and unlikely the final products after the chemical reaction, while they are perhaps good starting points to promote the chemical reaction. On the other hand, we found that Sr$_2$NiO$_2$Cl$_2$ + SrNiO$_3$ and SrNiO$_3$ + SrNiO$_2$ + SrCl$_2$ have similar total energies to that of the target compound Sr$_3$Ni$_2$O$_5$Cl$_2$.
This result suggests that, even by starting from the relatively unstable compounds including SrO$_2$, which was used in synthesis of similar oxychlorides, Sr$_2$NiO$_2$Cl$_2$~\cite{Tsujimoto},  Sr$_2$CoO$_3$Cl, and Sr$_3$Co$_2$O$_5$Cl$_2$~\cite{Cava}, a synthesis can be trapped in the middle of the reaction.

\begin{figure}
\begin{center}
\includegraphics[width=5.5 cm]{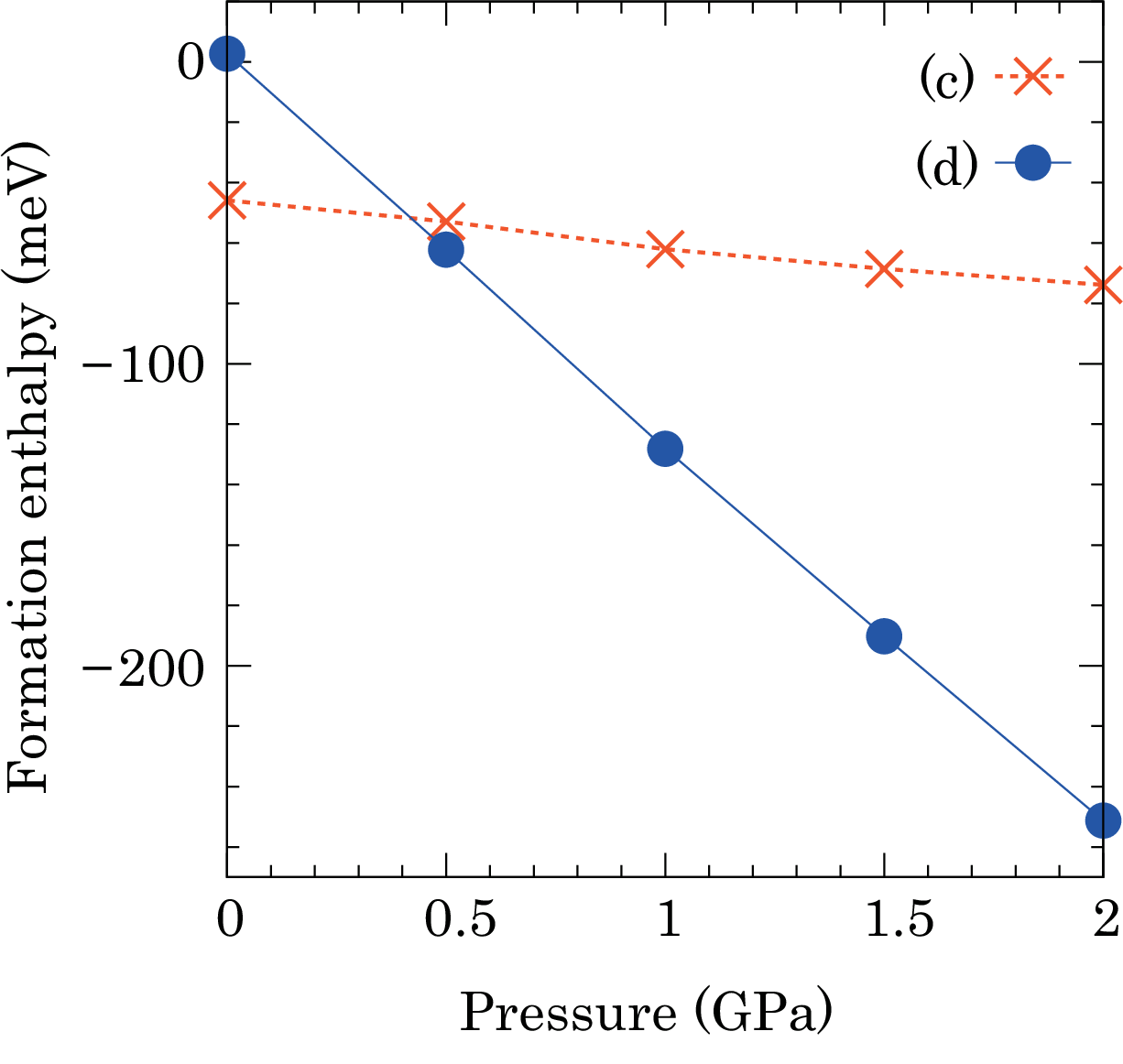}
\caption{Formation enthalpy of chemical reactions (c)(d) defined in Table~\ref{tab:form}.}
\label{fig:enthalpy}
\end{center}
\end{figure}

However, this problem might be resolved by the high-pressure synthesis, as adopted in Ref.~\onlinecite{Tsujimoto} for Sr$_2$NiO$_2$Cl$_2$. Figure~\ref{fig:enthalpy} presents that the formation enthalpy with respect to the applied pressure for chemical reactions (c)(d) that are defined in Table~\ref{tab:form}. Here, the enthalpy is defined as the sum of the total energy and the $pV$ term, where $p$ and $V$ are the pressure and the volume, respectively.
We found that the formation enthalpy goes to negative even for reaction (d).

Therefore, we can expect that the material we proposed in this study, Sr$_3$Ni$_2$O$_5$Cl$_2$, can be synthesized with the help of the high-pressure synthesis.
Our calculation also suggests that, under sufficiently high pressure, Sr$_2$NiO$_2$Cl$_2$ can appear as impurity phases because the formation enthalpy for chemical reaction (c) is not effectively lowered by pressure as shown in Fig.~\ref{fig:enthalpy}.
This result suggests that Sr$_2$NiO$_2$Cl$_2$ is possibly included as an impurity phase in the high-pressure synthesis of Sr$_3$Ni$_2$O$_5$Cl$_2$ under some conditions.

\begin{table}
\caption{Formation energies of Sr$_3$Ni$_2$O$_5$Cl$_2$.}
\label{tab:form}
\centering
\begin{tabular}{cccc}
\hline
\hline
& & (meV)\\
 \hline
(a) & 2SrO$_2$ + SrCl$_2$ + 2Ni + $\frac{1}{2}$O$_2$ $\to$ Sr$_3$Ni$_2$O$_5$Cl$_2$ & $-$5223 \\
(b) &  SrO$_2$ + SrO + SrCl$_2$ + 2NiO $\to$ Sr$_3$Ni$_2$O$_5$Cl$_2$& $-$2523\\
(c) & Sr$_2$NiO$_2$Cl$_2$ + SrNiO$_3$ $\to$ Sr$_3$Ni$_2$O$_5$Cl$_2$& $-$46 \\
(d) & SrNiO$_3$ + SrNiO$_2$ + SrCl$_2$ $\to$ Sr$_3$Ni$_2$O$_5$Cl$_2$& 3 \\
\hline
\hline
\end{tabular}
\end{table}

\section{Summary}

We have theoretically investigated the electronic structure of Sr$_3$Ni$_2$O$_5$Cl$_2$ and its structural stability.
Our phonon calculation has shown that this compound with the $I4/mmm$ tetragonal structure is dynamically stable, i.e., does not have imaginary-frequency phonon modes, even at ambient pressure.
The Ni-$d_{3z^2-r^2}$ orbital energy is pushed down by the crystal field where a half of apical O$^{2-}$ ions is replaced with a Cl$^{-}$ ion.
By this effect, the Ni-$d_{3z^2-r^2}$ orbital becomes rather closer to the half-filling in Sr$_3$Ni$_2$O$_5$Cl$_2$ than La$_3$Ni$_2$O$_7$ whereas the stoichiometric Ni-$3d$ band filling are decreased in
Sr$_3$Ni$_2$O$_5$Cl$_2$ ($d^7$) compared with La$_3$Ni$_2$O$_7$ ($d^{7.5}$).

By investigating superconductivity with the multi-orbital Hubbard model using the FLEX approximation combined with the linearized Eliashberg equation,
we have found two features originating from the characteristic crystal field.
First, interorbital interactions between the $t_{2g}$ and $e_g$ orbitals are detrimental for superconductivity.
On the other hand, a nearly half-filled Ni-$d_{3z^2-r^2}$ orbital enhances superconductivity.
In total, we have found that Sr$_3$Ni$_2$O$_5$Cl$_2$ is a promising new candidate of bilayer-nickelate superconductors, which can possess even higher $T_c$ than pressurized La$_3$Ni$_2$O$_7$, at ambient pressure. By calculating the formation enthalpy, we have found that the high-pressure synthesis can be a good way to actually produce Sr$_3$Ni$_2$O$_5$Cl$_2$.

It is noteworthy that the bilayer band structure consisting of Ni-$d_{3z^2-r^2}$ orbitals has a quite small bandwidth with a large bilayer splitting energy in Sr$_3$Ni$_2$O$_5$Cl$_2$. 
Namely, the present system perhaps lies in the strong-coupling regime of the bilayer square lattice, where the interlayer superexchange interaction $J_{\perp}$ stronger than the intralayer hopping amplitudes can enhance the superconductivity~\cite{DagottoScalapino}. This aspect can already be seen in La$_3$Ni$_2$O$_7$, but is more pronounced in the present system.
Thus, it is an intriguing future problem to investigate the present system from a strong-coupling picture.

Our study offers an important knowledge for materials design of bilayer-nickelate superconductors and will stimulate experimental investigation of mixed-anion bilayer-nickelate superconductors.

\acknowledgments
We thank Yoshihiko Takano, Hiroya Sakurai, and Kazuki Yamane for fruitful discussions.
This study was supported by JSPS KAKENHI Grant No.~JP22K03512, JP22K04907, and JP24K01333.
Part of computation was performed using the facilities of the Supercomputer Center, the Institute for Solid State Physics, the University of Tokyo, Japan, and Academic Center for Computing and Media Studies, Kyoto University, Japan.
 
\section*{Appendix A: Phonon band dispersion for La$_3$Ni$_2$O$_7$}

\begin{figure}
\begin{center}
\includegraphics[width=8 cm]{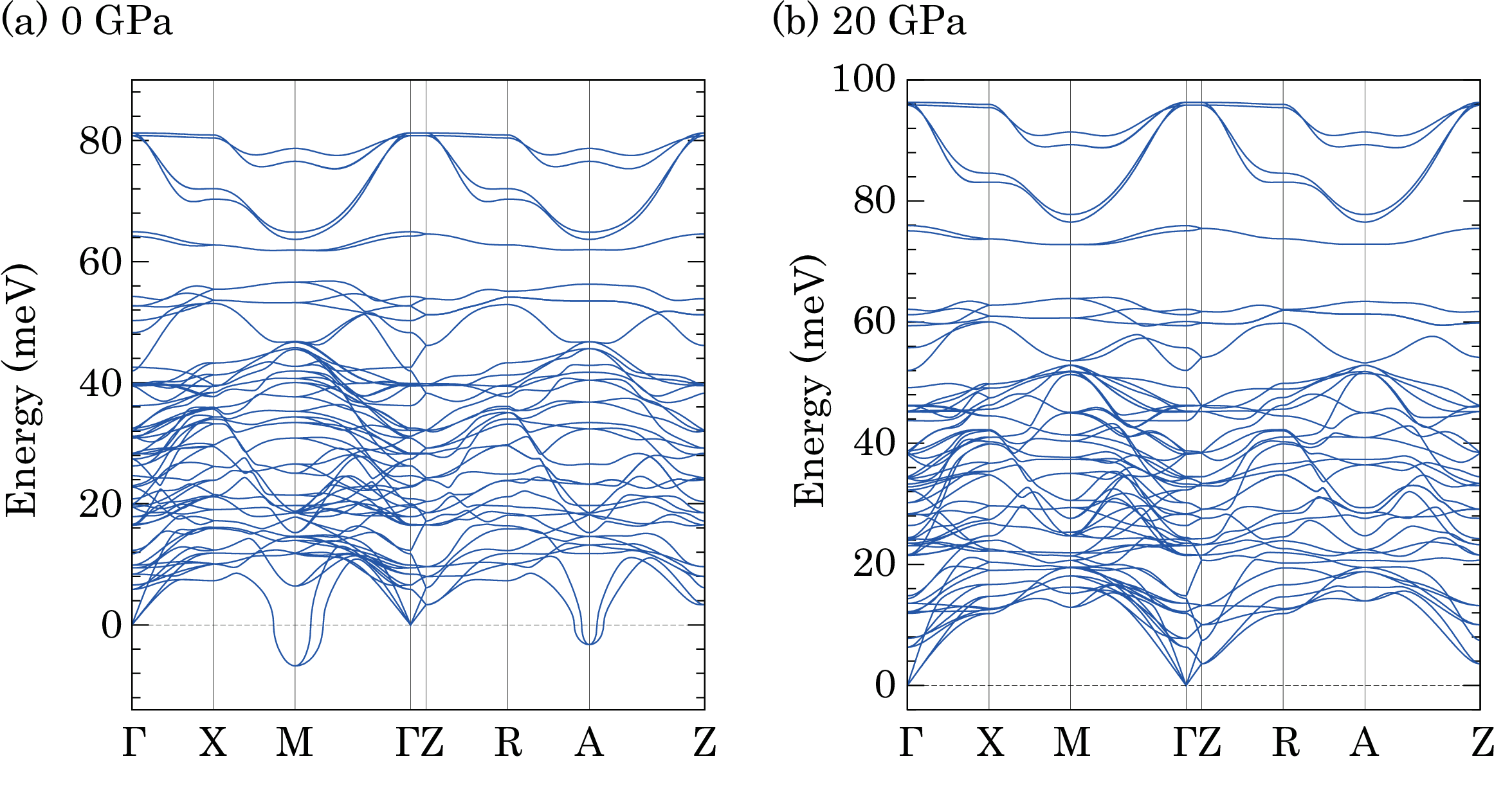}
\caption{Phonon band dispersion for La$_3$Ni$_2$O$_7$ at (a) 0 GPa and (b) 20 GPa, respectively.}
\label{fig:327_phonon}
\end{center}
\end{figure}

Figure~\ref{fig:327_phonon} presents the calculated phonon dispersion for La$_3$Ni$_2$O$_7$ under 0 and 20 GPa.
We confirmed that the crystal structure with the $I4/mmm$ space group is unstable without applying the pressure but is stable by applying a sufficient pressure as reported in several studies~\cite{Geisler_Hamlin,Wang_Li,Wang_Wang_50}.
 
\section*{Appendix B: Phonon band dispersion for Sr$_3$Ni$_2$O$_4$Cl$_3$}

\begin{figure}
\begin{center}
\includegraphics[width=8 cm]{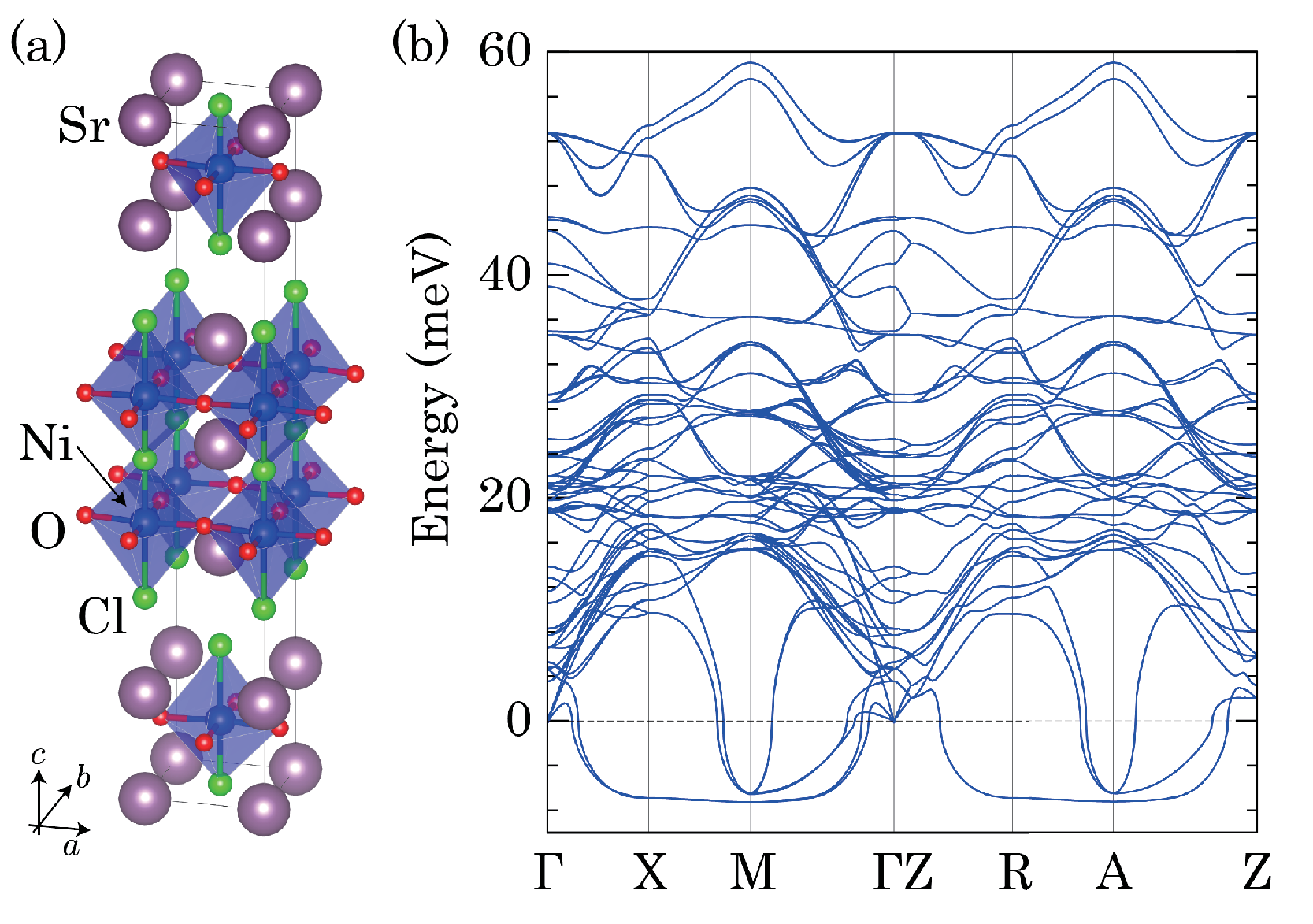}
\caption{(a) Crystal structure and (b) phonon band dispersion for Sr$_3$Ni$_2$O$_4$Cl$_3$.}
\label{fig:3243}
\end{center}
\end{figure}

Figure~\ref{fig:3243} presents the crystal structure and the calculated phonon dispersion for Sr$_3$Ni$_2$O$_4$Cl$_3$, where all the apical oxygens were replaced with chlorines.
We found that the crystal structure with the $I4/mmm$ space group is unstable for this material because imaginary-frequency phonon modes appear in Fig.~\ref{fig:3243}(b).

\section*{Appendix C: Electronic and phonon band structures for La$_{0.3}$Sr$_{2.7}$Ni$_2$O$_5$Cl$_2$}

\begin{figure}
\begin{center}
\includegraphics[width=8.5 cm]{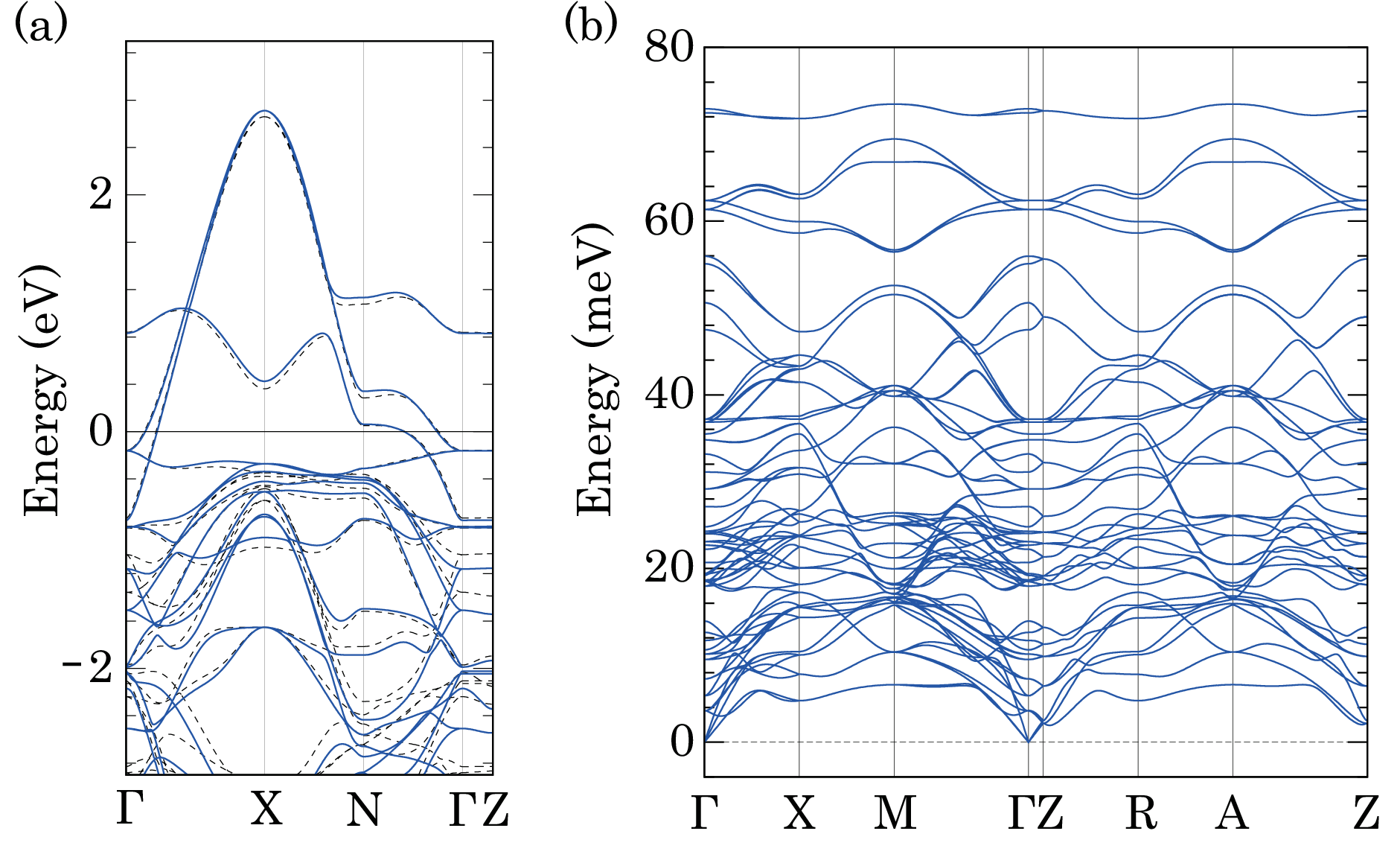}
\caption{(a) Electronic band structure for La$_{0.3}$Sr$_{2.7}$Ni$_2$O$_5$Cl$_2$ (blue solid lines) shown together with that for Sr$_{3}$Ni$_2$O$_5$Cl$_2$ with a shifted Fermi energy (black broken lines). (b) Phonon band dispersion for La$_{0.3}$Sr$_{2.7}$Ni$_2$O$_5$Cl$_2$.}
\label{fig:La10per}
\end{center}
\end{figure}

Figure~\ref{fig:La10per} presents the electronic band structure and the phonon dispersion for La$_{0.3}$Sr$_{2.7}$Ni$_2$O$_5$Cl$_2$, where the La doping was handled with the virtual crystal approximation.
We found that the rigid-band approximation adopted in this study was justified in Fig.~\ref{fig:La10per}(a) because the electronic band dispersion for La$_{0.3}$Sr$_{2.7}$Ni$_2$O$_5$Cl$_2$ agrees well with that for Sr$_{3}$Ni$_2$O$_5$Cl$_2$ with a shifted Fermi energy.
In addition, the absence of the imaginary-frequency phonon modes in Fig.~\ref{fig:La10per}(b) suggests that the stability of the $I4/mmm$ structure is kept against the La doping.

\section*{Appendix D: cRPA interaction parameters for Sr$_3$Ni$_2$O$_5$Cl$_2$}  
We evaluated the onsite interaction parameters for the ten-orbital model of Sr$_3$Ni$_2$O$_5$Cl$_2$ using the cRPA~\cite{PhysRevB.70.195104} with the projector method~\cite{projector_cRPA}.
For this purpose, we used the plane-wave cutoff energy of 450 eV, 200 bands including occupied and unoccupied states, and an $8\times 8\times 8$ ${\bm k}$-mesh. Obtained interaction parameters are shown in Table~\ref{tab:cRPA}.

\begin{table}[!h]
\caption{
Interaction parameters of onsite Coulomb repulsion $U,U'$ and Hund's coupling (pair-hopping) $J$ ($J'$) evaluated with cRPA.
The orbital indices $l,m=1,2,3,4,5$ indicate the $d_{x^2-y^2}, d_{3z^2-r^2}, d_{xy}, d_{yz}, d_{zx}$ orbitals, respectively.
For simplicity, we abbreviate four-orbital-index-representation of partially screened interaction integrals $V$,
namely, $V_{llmm}\rightarrow U'_{lm}$ and $V_{lmlm}(=V_{lmml})\rightarrow J_{lm}$.
Note that $U'_{lm}=U'_{ml}$, $J_{lm}=J_{ml}$, and $J_{lm}=J'_{lm}$. All in eV. \label{tab:cRPA}
}
\begin{tabular}{ p{10pt}  p{30pt}  | p{20pt} p{30pt}  p{30pt}} \hline\hline
&$l,m$ &  & $U,U'$ & $J,J'$  \\\hline
&$1,1$  & & 3.936 & \hspace{5pt}--- \\
&$1,2 $ & & 2.866 & 0.563\\ 
&$1,3$  & & 3.151 & 0.249\\
&$1,4$  & & 2.990 & 0.473\\
&$1,5$  & & 2.990 & 0.473\\
&$2,2$  & & 4.080 & \hspace{5pt}---\\
&$2,3$  & & 2.683 & 0.460\\
&$2,4$  & & 3.365 & 0.347\\ 
&$2,5$  & & 3.365 & 0.347\\
&$3,3$  & & 3.393 & \hspace{5pt}---\\
&$3,4$  & & 2.802 & 0.420 \\ 
&$3,5$  & & 2.802 & 0.420 \\
&$4,4$  & & 4.088 & \hspace{5pt}--- \\
&$4,5$  & & 3.040 & 0.472 \\
&$5,5$  & & 4.088 & \hspace{5pt}---
\\\hline\hline
\end{tabular}
\end{table}
 
\bibliography{3252}

\end{document}